\documentclass[useAMS,usenatbib]{mn2e}
\usepackage{latexsym}
\usepackage{amssymb}
\usepackage[dvips]{graphicx}
\usepackage{mathrsfs}

\title[The NIR to UV continuum of RLQs vs.~RQQs]{The NIR to UV continuum of radio loud vs.~radio quiet quasars}
\author[M. Labita, A. Treves and R. Falomo]{M. Labita$^{1}$\thanks{E-mail:
marzia.labita@uninsubria.it}, A. Treves$^{1}$ and R. Falomo$^{2}$\\
$^{1}$Department of Physics and Mathematics, University of Insubria, Via Valleggio 11, I-22100 Como, Italy\\
$^{2}$INAF, Astronomical Observatory of Padova, Vicolo dell'Osservatorio 5, I-35122 Padova, Italy}
\begin{document}

\date{Accepted ... Received ...; in original form ...}

\pagerange{\pageref{firstpage}--\pageref{lastpage}} \pubyear{0000}

\maketitle

\label{firstpage}

\begin{abstract}
Starting from a sample of SDSS quasars appearing also in the 2MASS survey, we study the continuum properties of  $\sim 1000$ objects observed in 8 bands, from NIR to UV. We construct the mean spectral energy distribution (SED) and compare and contrast the continua of radio loud (RLQ) and radio quiet (RQQ) objects. The SEDs of the two populations are significantly different, in the sense that RLQs are redder with power law spectral indices $\langle\alpha_{\rm RLQ}\rangle=-0.55\pm0.04$ and $\langle\alpha_{\rm RQQ}\rangle=-0.31\pm0.01$ in the spectral range between $10^{14.5}$ and $10^{15.35}$ Hz. This difference is discussed in terms of different extinctions, different disc temperatures, or slopes of the non-thermal component.%can be interpreted in terms of dust extinction or can be attributed to different temperatures of the accretion disks, in the sense that RLQs are on average more massive and then support cooler disks.  
\end{abstract}

\begin{keywords}
galaxies: active -- galaxies: nuclei -- quasars: general.
\end{keywords}

\section{Introduction}

%Several studies of quasar spectral energy distributions (SEDs) have <<been published, and 
A substantial effort has been dedicated to construct the spectral energy distributions (SEDs) for sizeable samples of quasars over the whole accessible range of the electromagnetic spectrum %from radio to hard X-rays 
(see e.g.~Sanders et al.~1989; Francis et al.~1991; Elvis et al.~1994; Richards et al.~2006). %(and $\gamma$-rays)
However, only a few papers focus on the comparison between the SEDs of radio loud (RLQ) and radio quiet (RQQ) quasars. Elvis et al.~1994 propose an overall spectrum from a sample of 47 quasars, divided in RLQs and RQQs, which shows that no distinction between the SEDs of the two subsamples is apparent in the range $100\mu - 1000$\AA. As noted by these authors, the considered sample is biased towards X-ray and optically bright quasars. Some indication of a possible difference between the NIR to optical colors of RLQs and RQQs in the 2MASS catalogue was reported by Barkhouse \& Hall (2001). Francis,
Whiting \& Webster (2000) found that the optical--NIR continuum is significantly redder in radio selected RLQs from the PKS Half-Jansky Flat-Spectrum survey than in optically selected RQQs from the Large Bright Quasar Survey (LBQS). These and other works (e.g.~Kotilainen et al.~2007) indicate that radio loud objects are possibly redder than their radio quiet counterparts, but the samples seriously suffer of selection effects against red radio quiet quasar, because RQQs are mostly collected from optical selected samples (e.g.~the LBQS, the first selection criterion of which is ``blue color'' of candidates; see Hooper, Impey \& Foltz 1997). 
White et al. suggest that redder quasars from the Sloan Digital Sky Survey (SDSS) are likely to be more radio-powerful than bluer objects, and this is an indication that the red color of radio loud objects is not completely due to the bias against red RQQs in optically selected samples, since the two classes of objects derive from the same survey.

Because of the importance of the distinction between RLQs and RQQs in the Unified Models of AGN, these results suggest and motivate a study aimed at investigating the properties of the continuum emission of quasars in the UV to NIR region, in order to compare and contrast the SEDs of radio loud and radio quiet objects.  %********************************The first step is the selection of a large quasar sample which is minimally biassed against the radio properties and the nuclear color of the objects: therefore we start our analysis from the SDSS quasar sample, which should be complete in this sense and extended enough. We also require that the selected objects have a 2MASS detection, in order to enlarge the spectral range towards the near-infrared. %From the observed fluxes in 8 bands ($u$, $g$, $r$, $i$, $z$, $J$, $H$, $K$) we construct the mean SED of about 1000 quasars, 12\% of which are radio loud. 

In Section \ref{secsample} we focus on the quasar sample selection criteria and on the host galaxy contribution. In section \ref{SED} we consider the SED construction and discuss the spectral shape. %features and we give a parametrization of the observed region. 
The RLQ and RQQ SEDs are then compared and contrasted. In the last section we provide a summary and a discussion of the results. 

%The aim of this paper is to compare and contrast 
%Nowadays there is wide consensus that ...

%The aim of this paper is to ...
%there is increasing evidence that ...
%This is consistent with a picture where...  

%In the next section we focus on the quasar sample selection criteria (\ref{secsample}). 
%In section \ref{secmas} we describe ... (\ref{secpho}--\ref{secvir}) and report on the results (\ref{seccomp}). This analysis .... The issue of ... is further discussed in paragraph \ref{calcolof}. In section \ref{secsum} our results are compared with those of literature and discussed within ...

Throughout this paper, we adopt a concordant cosmology with $H_0=70$ ~km~s$^{-1}$~Mpc$^{-1}$, $\Omega_m=0.3$ and $\Omega_{\Lambda}=0.7$. 

\section{Sample selection}
\label{secsample}
%%%%%%%%%%%%%%%%%%%%%%%%%%%%%%%%%%It has been suggested (see for example Francis et al.) that the observed color difference between RLQs and RQQs is partially due to the common bias against the presence of red RQQs in most catalogues, as RQQs are often optically selected objects. 
In order to mark similarities and differences of the continuum spectroscopic properties of RLQs and RQQs, our first aim is the selection of a quasar sample which is minimally biassed against the radio properties and the nuclear color of the objects.
%(add che sloan contiene piccolo bias radio) %To minimize any possible bias against red RQQs, 
We start our analysis from the Sloan Digital Sky Survey (SDSS) quasar catalogue III (Schneider et al.~2005); we then require that objects are in the observation area of the Faint Images of the Radio Sky at Twenty-cm (FIRST, Becker et al.~2003), which is the reference catalogue for radio data, in order to allow the distinction between RLQs and RQQs. Finally, this sample is cross-correlated with the Two Micron All Sky Survey (2MASS, Cutri et al.~2003) in order to have detections in the near-infrared (NIR) bands.

\subsection{SDSS quasar catalogue}
The SDSS quasar catalogue (46420  objects) covers about 4188 deg$^{2}$ and selects objects that have luminosities  larger than $M_i=-22$, have at
least one emission line  with  FWHM  larger  than  1000  km$/$s  or  are
unambiguously  broad absorption line objects,  are fainter than $i=15.0$
and have highly reliable redshifts. 

Note that the SDSS quasar catalogue suffers from a small bias in favour of radio loud objects, as a FIRST detection is one of the starting criteria to select quasar candidates; anyway White et al.~(2007) verified that this introduces at most a very minor bias towards higher radio-optical ratios.

\subsection{FIRST: distinction between RLQs and RQQs}
Starting from the SDSS quasars, we focus on those which are in the FIRST observation area, but %(which are the 90\% of the whole SDSS quasar catalogue). %Again, no bias against the color or the radio power is introduced. 
we don't exclude {\it a priori} the objects which are under the FIRST flux limit (1 mJy), so that no bias against the radio power is introduced. %In Fig.~\ref{sloan_first} we show the absolute magnitude $M_i$ vs. redshift for the SLOAN quasar in the FIRST field.

The distinction between radio loud and radio quiet objects is usually made referring to the ratio between the radio and optical flux. Here we assume that a QSO is radio loud if:
\begin{equation}\label{LQ}
R_{\rm r-o}=\frac{F_{\nu}(1.5\cdot10^{9}\textrm{Hz})}{F_{\nu}(6\cdot10^{14}\textrm{Hz})}>10
\end{equation}
and radio quiet otherwise\footnote{Equation (\ref{LQ}) should be applied to rest-frame data, but this correction requires the assumption of a spectral shape of the objects. Here we choose to evaluate this ratio based on observed data; we verified that the difference is completely negligible.} (e.g. Kellermann et al.~1989).  Note that our division between RLQs and RQQs should not be considered a physical
bimodality, an issue on which there is still controversy in
the literature (e.g.~Goldschmidt et al.~1999; Ivezi\'c et al.~2002; Jiang at al.~2007). Yet our division is a simple separation in two groups of high or low radio
power with respect to a certain limit. We verified that all the results presented here remain substantially unchanged adopting different limits (5 or 30) for the distinction between RL and RQ objects.

There are not radio data for all the objects in the sample (91\% of the objects are below the FIRST radio flux limit), and obviously the larger part of these are expected to be radio quiet objects. In order not to introduce a bias against the ratio of RLQs and RQQs, we select all the objects which have $g<18.9$, so that for this subsample we can discriminate between RLQs and RQQs. In fact, if an object has a radio flux under the FIRST limit (i.e. $F_{\nu}(1.5\cdot10^{9}\textrm{Hz})<1$mJy) and $g<18.9$  (i.e. $F_{\nu}(6\cdot10^{14}\textrm{Hz})>0.1$mJy), because of condition (\ref{LQ}) it is a RQQ, while we could not discriminate if $g>18.9$. This restriction reduces the sample by 66\%, %but this procedure allows to 
keeping the original ratio between the number of RQQs and RLQs. We obtain a sample 
%(hereafter SLOAN--FIRST sample) 
of 14395 objects, 1105 of which are RLQs and 13290 are RQQs%
%\footnote{Condition \ref{LQ} should be applied to rest-frame data, but this correction requires the assumption of a spectral shape of the objects. Here we choose to evaluate this ratio based on observed data; we verified that the mismatch is completely negligible.}
. From the SLOAN data, we derived the $u$, $g$, $r$, $i$, $z$ magnitudes. The values were corrected for Galactic extinction, following the indications given in Schneider et al.~(2005). %but no k-correction was applied because it would have required an assumption {\it a priori} on the SEDs of RLQs and RQQs. %For the same reason, the absolute magnitudes are not rest-frame unless otherwise specified.

\subsection{NIR detections from 2MASS}\label{2mass}
The SLOAN--FIRST sample was then cross-correlated with the 2MASS catalogue, in order to obtain the $J$, $H$ and $K$ magnitudes of the objects. The 2MASS survey is a collection of near-infrared uniformly-calibrated observations of the entire sky; sources brighter than about 1 mJy in each band were detected with a signal-to-noise ratio greater than 10. The cross-correlation strongly reduces  the sample, which now consists of 1761 common objects  with observations in 8 bands each. %, about the 14\% of the original SLOAN--FIRST sample. 
We assumed two objects are matched if their separation angle is less than 0.2 arcsec, which roughly corresponds to the position error of the catalogues\footnote{We verified that this limit on the separation angle causes the omission of about 20\% objects which belong to both catalogues, but drastically reduces the probability of mismatch in the selection of the objects.}.  Obviously the request of a 2MASS detection introduces a bias in the sample, as
among high redshift (low luminosity) objects the redder ones are more likely
detected. In the following we will come back on this issue. %, showing that this bias can be seen {\it a posteriori} as a confirmation of our hints. 

\subsection{Host galaxy component}
A host galaxy component could affect the luminosity measures in part of the selected quasars. We require that the selected objects to have a negligible galaxy component with respect to the nuclear flux. Assuming that all the galaxies are ellipticals (SED from Mannucci et al.~2001), purely passively evolving (Bressan, Chiosi \& Fagotto~1994), with $M_R=-22.6$ for RLQs and $M_R=-21.8$ for RQQs at $z\sim0$ (see for example Kotilainen et al.~2006), we give a first order estimate of the galaxy component flux in the 8 SLOAN \& 2MASS filters. An object is excluded from the sample if the estimated flux of the host galaxy is greater than 20\% of the observed flux%
%\footnote{We exclude the objects instead of correcting the flux for the star-light contamination as our estimate of the galaxy component is on average good, but quite rough on the single object.}
. The excluded objects ($\sim1000$) are mostly weak, nearby ($z<0.5$) quasars.
 We also tested if the removal of quasars with substantial host galaxies could be done by looking at the
   difference between PSF and model magnitudes in SDSS. The results reported in the following remain substantially unchanged.%The agreement between the two procedure is quite good at
%high redshift. ALTRO?????}
\\ % (or, likely, Seyfert objects which have been classified as quasars in the SLOAN catalogue).\\

\subsection{Final sample}
%\subsection{Redshift and magnitude distributions matching}\label{SEC_matched_sample}
The final sample (see Table \ref{elec2}, available in complete form electronically) consists of 887 QSOs, of which 774 RQQs and 113 RLQs.
% \begin{enumerate}
%\item SLOAN quasar catalogue, 46420 objects
%\item FIRST field of observation, 41737 objects (90\% of the SLOAN sample)
%\item $g<18.8$ in order to allow the distinction between RLQs and RQQs, 14395 objects, 8\% of which are RLQ (31\% of the SLOAN sample)
%\item 2MASS detection, 1761 %2015 
%objects, 11\% of which are RLQ (4\% of the SLOAN sample)
%\item negligible host galaxy component, 887 %934 
%objects, 13\% of which are RLQ (2\% of the SLOAN sample).
%\end{enumerate}
%Table \ref{elec2} (completely available in electronic form) contains 

\begin{table*}
 \centering
 \begin{minipage}{\textwidth}
  \centering
  \caption{The final sample. Monochromatic fluxes are derived from SDSS (col.~4--8), 2MASS (col.~9--11) and FIRST (col.~12) and corrected for Galactic extinction; units are $10^{-4}$Jy. In col.~(12), ``n.d.''~means ``not detected''. The complete version of this table is available in electronic form.}\label{elec2}
\begin{tabular}{@{}ccc|ccccc|ccc|c|cc@{}}
\hline
\hline
&&&&&SDSS&&&&2MASS&&FIRST&&\\
$\alpha$ (J2000)&	$\delta$ (J2000)&	   $z$&	   $F_u$&      $F_g$&      $F_r$&    $F_i$&	  $F_z$&       $F_J$&	$F_H$&$F_K$&   $F_{\rm 20cm}$& 
$R_{\rm r-o}$\\
(1)&(2)&(3)&(4)&(5)&(6)&(7)&(8)&(9)&(10)&(11)&(12)&(13)\\
%$[$h m s$]$&	$[$d m s$]$&   	&   $[10^{-4}{\rm Jy}]$&       $[10^{-4}{\rm Jy}]$&       $[10^{-4}{\rm Jy}]$&      $[10^{-4}{\rm Jy}]$&  $[10^{-4}{\rm Jy}]$&   
%$[10^{-4}{\rm Jy}]$& $[10^{-4}{\rm Jy}]$&  $[10^{-4}{\rm Jy}]$& $[10^{-3}{\rm Jy}]$&     \\  
\hline
00 01 10.97&  -10 52 47.5&  0.528& 2.69& 3.40 & 3.21  & 3.53 & 3.24 & 3.38 & 4.82 & 6.46&  n.d. & $<10$\\
00 14 20.37&  -09 18 49.4&  1.083& 2.01& 2.82 & 4.25  & 3.95 & 3.97 & 5.26 & 4.12 & 6.80& 57.3  & 19.6\\
00 18 55.22&  -09 13 51.2&  0.756& 2.86& 4.25 & 5.09  & 5.12 & 5.62 & 5.88 & 6.07 & 7.12&  n.d. & $<10$\\
00 24 11.66&  -00 43 48.1&  1.795& 1.73& 2.08 & 2.29  & 2.69 & 2.57 & 2.67 & 3.44 & 4.88& 11.7  & 5.55\\
00 29 14.21&  -09 00 16.1&  2.091& 0.71& 1.16 & 1.66  & 2.02 & 2.70 & 3.60 & 3.46 & 5.37&  n.d. & $<10$\\
00 30 09.40&  -09 02 23.1&  1.786& 2.59& 2.51 & 2.69  & 3.42 & 3.41 & 3.09 & 5.26 & 3.25&  n.d. & $<10$\\
00 34 13.04&  -01 00 26.9&  1.292& 4.72& 5.03 & 5.64  & 5.81 & 5.54 & 5.08 & 5.62 & 6.86&  n.d. & $<10$\\
00 36 26.11&  -09 00 14.2&  0.951& 1.86& 1.93 & 2.05  & 1.89 & 1.89 & 2.88 & 1.83 & 4.98&  n.d. & $<10$\\
00 36 33.93&  -10 12 28.8&  2.082& 1.63& 1.89 & 2.06  & 2.33 & 2.68 & 2.69 & 2.79 & 4.05&  n.d. & $<10$\\
00 37 14.82&  -00 45 54.1&  1.020& 1.66& 1.86 & 2.12  & 1.96 & 2.16 & 2.48 & 1.65 & 3.41&  n.d. & $<10$\\
00 38 23.81&  -00 00 25.1&  1.605& 1.63& 1.79 & 2.12  & 2.52 & 2.67 & 3.33 & 3.12 & 3.11&  n.d. & $<10$\\
00 38 42.66&  -09 47 12.8&  1.989& 3.65& 3.97 & 4.45  & 4.82 & 5.29 & 4.39 & 3.60 & 5.18&  n.d. & $<10$\\
00 42 22.29&  -10 37 43.8&  0.424& 9.13& 11.3 & 9.75  & 10.0 & 12.5 & 7.87 & 10.8 & 17.8&  n.d. & $<10$\\
00 46 13.54&  +01 04 25.7&  2.152& 1.51& 1.90 & 2.27  & 2.70 & 3.11 & 4.02 & 4.41 & 6.35& 30.4  & 15.6\\
\ldots&\ldots&\ldots&\ldots&\ldots&\ldots&\ldots&\ldots&\ldots&\ldots& \ldots&\ldots&\ldots&\\
\hline
\end{tabular}
\end{minipage}
\end{table*}

\begin{figure}
\includegraphics[width=0.45\textwidth]{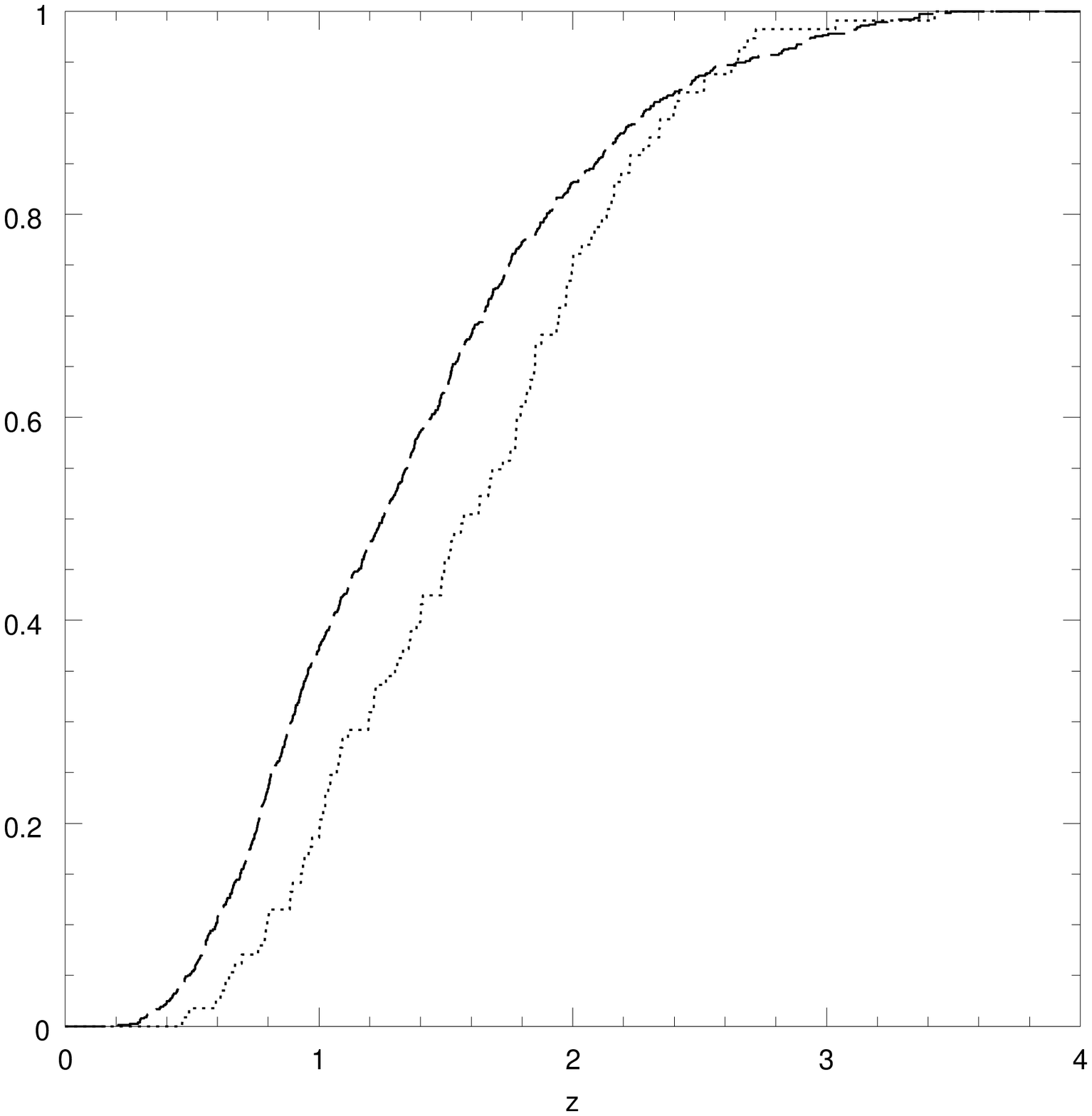}
\caption[]{Cumulative redshift distribution of the final 
%SLOAN-2MASS 
sample. Dotted line refers to the RLQ and dashed line to the RQQ. }
\label{zdistrib}
\end{figure}

\begin{figure}
\includegraphics[width=0.45\textwidth]{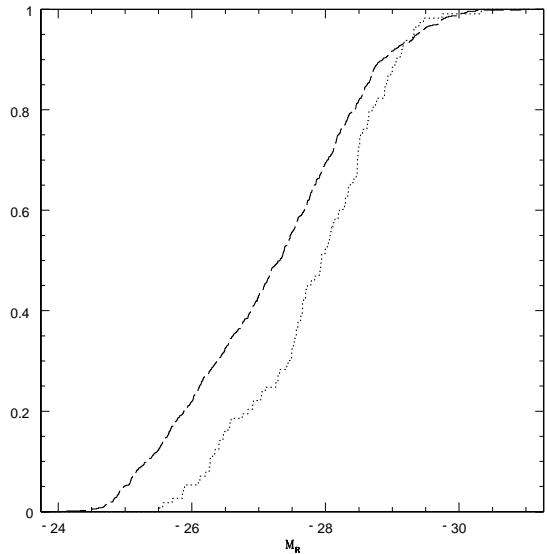}
\caption[]{Cumulative $M_R$ distribution of the final 
%SLOAN-2MASS 
sample. Dotted line refers to the RLQ and dashed line to the RQQ.% $M_R$ is evaluated from the nearest observed magnitudes assuming a spectral template from Francis et al.
}
\label{Mdistrib}
\end{figure}

Figures \ref{zdistrib} and \ref{Mdistrib} show the magnitude and redshift distributions of the radio loud and radio quiet subsamples, indicating a substantial difference between the two classes. %The distributions in redshift and optical magnitude of RQQs are significantly different from those of RLQs. 
This behaviour is well known and it is usually attributed to the cosmological evolution of quasar radio properties or to a different density evolution for RLQs and RQQs (e.g.~White et al.~2007; Jiang et al.~2007). %In section (\ref{SED_bias}) we show that our results are independent on this behaviour.
%NEW, REVERSE
%
The issue is obviously relevant to our discussion and it is considered in section \ref{SED}. %, in order to compare the shape of the SEDs of RLQs and RQQs, we will extract randomly from the RQQ sample a number of populations (``RQQ matched samples'') that are well matched in the redshift and magnitude distributions with the RLQs. Figure \ref{matched_distrib} shows the distributions of one RQQ matched sample compared to the distributions of the RLQ population. Hence, our results describe the shape of the SED in function of the radio power only, independently of any other observable parameter. 
%Since we are interested in the study of the shape of the SED in function of the radio power of the source, independently of any other parameter, we select from the radio quiet population a subsample of objects which have the same distributions in redshift and magnitude of the RLQs. 
%MATCHED SAMPLES!!!!!!!!!!!!!!!!!!!!!!!!!!!!!!!!!!!!!!!!!!!!!!!!!!!

%\begin{figure}
%\includegraphics[width=0.45\textwidth]{matched_distrib}
%\caption[]{FALLA CUMULATIVA! Redshift (left) and $M_R$ (right) distribution of the RLQ matched sample compared to the RLQ distributions. Note that the RLQ matched sample is a subsample of the RQQ population such that it has nearly the same distribution in redshift and magnitude of RLQs.}
%\label{matched_distrib}
%\end{figure}

%Hereafter, this RQQs matched sample will be referred to simply as ``RQQs''.  
%(TABELLA ELETTRONICA?)

\section{Spectral energy distributions (SEDs)}
\label{SED}
%\subsection{SED construction}
%\label{SED_constr}
In order to construct and compare the SEDs of RQQs and RLQs, we first evaluated the rest-frame SEDs of single objects. For each one, the SED consists of 8 data-points $\log(\nu)$--$\log(\nu L_{\nu})$, which are linearly interpolated. %which have been inferred and put rest-frame from the observed fluxes in the 5 SLOAN filters ($u,\,g,\,r,\,i,\,z$), previously corrected for Galactic extinction, following the indications given in Schneider et al.~(2005), and three 2MASS filters ($J,\,H,\,K$). %Note that this procedure doesn't require any assumption on the spectral shape of the objects. 
%Following the usual procedure (e.g.~Elvis), 
In order not to be affected by the well known bias which favours more luminous objets at higher redshift, using the standard procedure (e.g.~Elvis et al.~1994), 
the SEDs  of the radio loud and radio quiet samples have then been normalized separately at $10^{14.8}$Hz. %: for each object, where the value of $\log(\nu 

%L_{\nu}(10^{14.8}\textrm{Hz}))$ has been obtained interpolating the 2 nearest data-points of the SED. %; linear extrapolation has been used beyond data-points. 
%This procedure has been applied separately to the radio loud and the radio quiet samples. 
The average SEDs of the two samples have been determined evaluating the average value of $\log(\nu L_{\nu})$ at each frequency.

%\begin{figure}
%\includegraphics[width=0.45\textwidth]{sedtot_controllaflussipesante02}
%\caption[]{SEDs of all the objects (red for RLQs and blue for RQQs) normalized at $\log(\nu L_{\nu}(10^{14.8}\textrm{Hz}))$ and  mean SEDs of RLQs (yellow) and RQQs (green).}
%\label{sed_total}
%\end{figure}

%\begin{figure}
%\includegraphics[width=0.45\textwidth]{sedpochipunti_controllaflussipesante02}
%\caption[]{(A ME NON PIACE MOLTO QUESTA FIGURA , LA SOSTITUIREI FORSE CON UNA A FARFALLA COME IN ELVIS... NON SO)Mean SEDs of RLQs (red, circles) and RQQs (blue, triangles) normalized at $\log(\nu L_{\nu}(10^{14.8}\textrm{Hz}))$. We show only the data-points which are not contaminated by strong emission lines. The error bars represent the standard deviation on the data-points.}
%\label{sed_pochipunti}
%\end{figure}

%\begin{figure}
%\includegraphics[width=0.45\textwidth]{solosed_controllaflussipesante02}
%\caption[]{Mean SEDs of RLQs (red) and RQQs (blue) normalized at $\log(\nu L_{\nu}(10^{14.8}\textrm{Hz}))=0$. Solid lines: mean SED; dashed lines: standard deviation; dotted lines: errors on the mean.}
%\label{mean_sed_relative}
%\end{figure}

\begin{figure}
\includegraphics[width=0.45\textwidth]{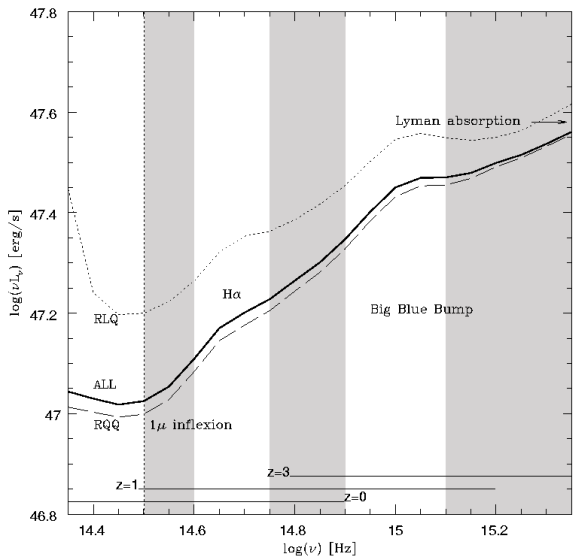}
\caption[]{SED for the entire sample of quasar (solid line), 
%normalized at $\log(\nu L_{\nu}(10^{14.8}\textrm{Hz}))$. The principal features are indicated. The SED can be roughly divided in 3 regions described by different power laws, as indicates in figure. Mean SEDs of 
RLQs (dotted line) and RQQs (dashed line). The horizontal lines show the spectral region available at different redshifts. The grey regions indicate the spectral ranges where the power law fit is performed.}
\label{figall}
\end{figure}

\begin{figure}
\includegraphics[width=0.45\textwidth]{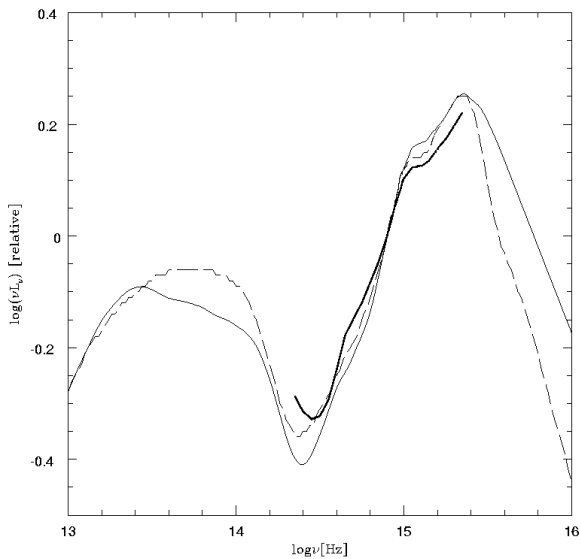}
\caption[]{SED for our entire sample of quasar (thick solid line), compared to the SEDs obtained by Elvis et al.~1994 (thin solid line) and by Richards et al.~2006 (thin dashed line). The three SEDs have been normalized at $10^{14.9}\,$Hz.}
\label{cfrelvisrich}
\end{figure}

Fig.~\ref{figall}
%\footnote{The frequency bins shown in all figures reflect the original separation between the filters in which the observations are made.} 
shows mean SED of the whole sample and those of the RLQ and RQQ subsamples, and in Table \ref{tab_mean_sed} we give the number of data-points which contribute to the mean for selected values of $\log(\nu)$, the corresponding average $z$ and the mean value of $\log(\nu L_{\nu})$ with corresponding errors. The overall SEDs are similar to those reported by Elvis et al.~1994 and Richards et al.~2006  (see Fig.~\ref{cfrelvisrich}).
%, its standard deviation and the error on the mean. In Figure \ref{sed_pochipunti} we show again the mean SED of RLQs and RQQs: only the data-points which are not contaminated by strong emission lines (...) are plotted. %Fig.~\ref{mean_sed_relative} shows the RLQs and RQQs rest-frame relative SEDs, normalized at $\log(\nu L_{\nu}(10^{14.8}\textrm{Hz}))=0$. The error on the mean SED and the standard deviation over all the data-points are also plotted. %(CHIARIRE LA DIFFERENZA TRA LE FIGURE O MODIFICARLE)
%\subsection{SED description}
%\label{SED_discuss}
%Fig.~\ref{mean_sed_relative} shows the relative mean SEDs of the radio loud and radio quiet subsamples. The shape of the two mean SEDs are qualitatively very similar, but the slope of the underlying power law is significantly different. This point will be discussed later on; 
%%%%%%%%%%%%%%%%%%%%Here we focus on the common features of the typical SEDs of quasars in general. 
%In Figure \ref{figall}, it is shown the composite SED of the whole sample of quasars. %The agreement with the composite SEDs of Elvis et al.~and Richards et al.~is good. 
%AGGIUNGERE CONSISTENZA CON LETTERATURA ELVIS-RICH
The strongest feature is the so-called 1 micron inflexion% ($1\mu\sim10^{14.5}{\rm Hz}$)
, where the slope of the SED changes sign in a $\nu L_{\nu}$ vs.~$\nu$ representation. Here, the optical-ultraviolet continuum rises above the infrared and forms a ``$UV$ bump'' (e.g.~Shields 1978; Malkan \& Sargent 1982; Elvis et al.~1994), %. Variability studies show that this is a separate component from the infrared (Cutri et al.~1985) since it varies much more strongly. 
which is most often interpreted in terms of thermal emission from an accretion disc (e.g., Malkan 1983; Czerny \& Elvis 1987).

At high frequencies ($\nu\gtrsim10^{15.35}{\rm Hz}$) the SEDs of all the objects are strongly contaminated by the Lyman $\alpha$ forest; the small bump at $\sim10^{14.66}{\rm Hz}$ is due to the presence of the H$\alpha$ large emission line.

%The composite SED is roughly divided in 3 regions described by 3 underlying power laws with different spectral indices ($\alpha_1=-1.17$, $\alpha_2=-0.05$, $\alpha_3=-0.83$; see Figure \ref{figall}). Since the so-called Big Blue Bump is often interpreted in terms of thermal emission from an accretion disk, we tried a fit with a black body function, but a single temperature is not sufficient to describe the emission. %For the purposes of this work, i.e.~ the comparison between the SEDs of RLQs and RQQs, we need to find a parameter 
%Note anyway that the range $10^{14.5}-10^{15.35}$ Hz is well described by a single power law. Hereafter, in the comparison between RLQs and RQQs, we choose to fit this region with one power law and discuss in terms of a single spectral index $\alpha$, because one parameter is much better constrained and allows an easier comparison between RLQs and RQQs.

%\begin{figure}
%\includegraphics[width=0.45\textwidth]{compositespectrum}
%\caption[]{Composite SED of the whole sample normalized at $\log(\nu L_{\nu}(10^{14.8}\textrm{Hz}))$. The principal features are indicated. The SED can be roughly divided in 3 regions described by different power laws, as indicates in figure.}
%\label{comp_spec}
%\end{figure}

%salita verso sinistra, righe di emissione, componente powerlaw..., bump termico... Lyman forest...

%\subsection{NIR--UV spectral index}
%\label{SED_alpha}

%The shape of the mean SEDs of RLQs and RQQs are qualitatively similar, but the slope is significantly different. 
It is apparent that RLQs are softer than RQQs. In particular, we compare the slope between the 1 micron inflexion and the Lyman $\alpha$ forest. The spectral index $\alpha$ (such that $F_{\nu}\propto \nu^{\alpha}$) is derived for each object by fitting with a power law the observed fluxes in the spectral range between $10^{14.5}$ and $10^{15.35}$ Hz, 
%From the SED of each object, we have 8 data-points ($\nu-\nu L_{\nu}$): the spectral index $\alpha$ (such that $F_{\nu}\propto \nu^{\alpha}$) is derived by fitting with a power law the data-points in the spectral range between $10^{14.5}$ and $10^{15.35}$ Hz%\footnote{Since the bump is often interpreted in terms of thermal emission from an accretion disk, we first tried a fit with a black body function, but a single temperature is not sufficient to describe the emission. As the considered spectral range is sufficiently well fitted by a power law, here we prefer to discuss in terms of spectral index, which is much better constrained and allows an easier comparison between RLQs and RQQs.}
%, i.e.~between the 1 micron inflexion and the Lyman forest. 
excluding the data points affected by the most prominent emission features (i.e. $\sim10^{14.6}-10^{14.75}$ Hz, $10^{14.9}-10^{15.1}$ Hz). 
%This procedure allows the determination of $\alpha$ for almost all the objects in the sample (887 of 934\footnote{In 47 cases, the exclusion of the spectral regions specified above left a single data-point or none at all per object.}): 

\begin{figure}
\includegraphics[width=0.45\textwidth]{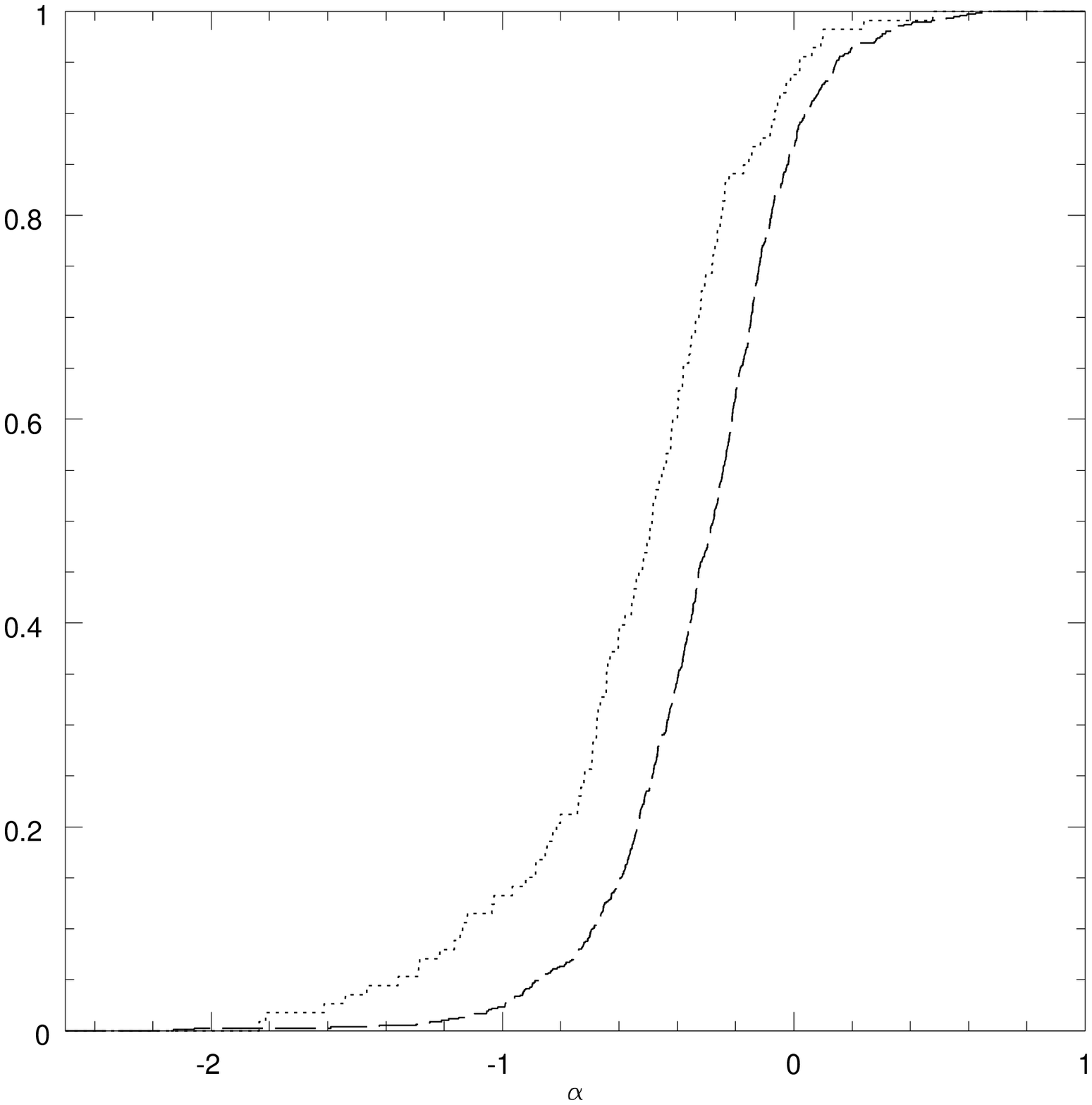}
\caption[]{Cumulative distribution of the spectral index for the RL (dotted line) and RQ (dashed line) subsamples.}
\label{alpha_distrib}
\end{figure}

Table \ref{pertantiz} reports the mean values of $\alpha$, the standard deviation and the error on the mean value for the RL and RQ subsamples and Figure \ref{alpha_distrib} shows the cumulative distribution of the spectral index for the two populations. %Note that the spectral range on which the fit is performed depends on the redshift of the objects, and so the spectral indices of different objects can be referred to different regions. Consequently, different spectral ranges have different weights on the average value, which is dominated by the optical region. This bias can be evaluated by fitting directly the mean SED of RLQs and RQQs, and we verified that no significant difference in the mean value of $\alpha$ is introduced, both for the RL and RQ subsamples. 
The slopes of the RLQs and the RQQs are statistically different at more than 99\% confidence level, with RLQs being softer. The spectral indices of the two populations in the optical region differ by $\sim0.2$, with $\alpha=-0.55\pm0.41\,(\pm0.04)$ for RLQs and $\alpha=-0.31\pm0.32\,(\pm0.01)$ for RQQs (where the uncertainties are the standard deviations and the errors on the mean respectively). A Kolmogorov-Smirnov test indicates that the probability of RLQs and RQQs to be drawn from the same underlying population is negligible ($P\sim 10^{-7}$). This is a strong indication that the two populations are intrinsically different, in the sense that radio loud objects are systematically redder than radio quiet quasars. 

 In Section \ref{2mass} we noted that the request of a 2MASS detection introduces a bias in the sample, as
among high redshift (low luminosity) objects the redder ones are more likely
detected. Consequentely, the blue side of the resulting SEDs (which is
dominated by high redshift objects) are slightly softer with respect to the
SED of an unbiassed sample. This effect can be quantified by constructing the
mean SEDs on the whole SLOAN sample, i.e. without the request of a 2MASS
detection. The shape of the SEDs is modified as expected by the bias introduced,
but it is noticeable that the difference between the SEDs of RLQs and RQQs
(which is our main result) remains substantially unchanged if we consider
the whole SLOAN sample to construct the SED for $\log(\nu)>14.8$ (where the 2MASS data are uninfluent) and the SLOAN--2MASS sample to construct the SED for $\log(\nu)<14.8$ (where the introduced bias is negligible). The average spectral indices of the RL and RQ populations, as determined on these SEDs, are respectively $\alpha=-0.40$ and $\alpha=-0.22$, with again $\Delta\alpha\sim0.2$.

\begin{table*}
 \centering
 \begin{minipage}{\textwidth}
  \centering
  \caption{For selected frequencies, we report the number of data-points which contribute to the SED of RLQs and RQQs, the mean redshift and its standard deviation, the luminosity, its standard deviation and the error on the mean. The luminosities are normalized at $\log(\nu L_{\nu}(10^{14.8}\textrm{Hz}))$.}\label{tab_mean_sed}
\begin{tabular}{@{}c|ccc|ccc@{}}
\hline
\hline
\phantom{.}&\phantom{.}&RLQ&\phantom{.}&\phantom{.}&RQQ&\phantom{.}\\
\hline
$\log(\nu)$&Num.&$z$&$\log(\nu L_{\nu})$&Num.&$z$&$\log(\nu L_{\nu})$\\
$[$Hz$]$&&&$[$erg s$^{-1}]$&&&$[$erg s$^{-1}]$\\
\hline
     14.35   &      3  &  $0.514  \pm 0.070$ & $47.446  \pm 0.292  (\pm0.207) $&     87   & $0.481  \pm 0.098$ & $47.013  \pm 0.132  (\pm0.014)$\\
     14.40   &     13  &  $0.671  \pm 0.115$ & $47.241  \pm 0.163  (\pm0.047) $&    190   & $0.613  \pm 0.145$ & $47.003  \pm 0.123  (\pm0.009)$\\
     14.45   &     26  &  $0.815  \pm 0.171$ & $47.197  \pm 0.172  (\pm0.034) $&    301   & $0.725  \pm 0.190$ & $46.993  \pm 0.117  (\pm0.007)$\\
     14.50   &     39  &  $0.924  \pm 0.213$ & $47.200  \pm 0.145  (\pm0.023) $&    398   & $0.829  \pm 0.250$ & $46.999  \pm 0.105  (\pm0.005)$\\
     14.55   &     55  &  $1.069  \pm 0.293$ & $47.223  \pm 0.131  (\pm0.018) $&    513   & $0.961  \pm 0.331$ & $47.028  \pm 0.102  (\pm0.004)$\\
     14.60   &     76  &  $1.256  \pm 0.396$ & $47.264  \pm 0.115  (\pm0.013) $&    609   & $1.077  \pm 0.407$ & $47.084  \pm 0.097  (\pm0.004)$\\
     14.65   &     95  &  $1.412  \pm 0.475$ & $47.321  \pm 0.091  (\pm0.009) $&    686   & $1.184  \pm 0.489$ & $47.145  \pm 0.089  (\pm0.003)$\\
     14.70   &    106  &  $1.511  \pm 0.536$ & $47.353  \pm 0.072  (\pm0.007) $&    733   & $1.261  \pm 0.559$ & $47.176  \pm 0.074  (\pm0.003)$\\
     14.75   &    112  &  $1.576  \pm 0.591$ & $47.363  \pm 0.040  (\pm0.004) $&    757   & $1.311  \pm 0.615$ & $47.205  \pm 0.039  (\pm0.001)$\\
     14.80   &    113  &  $1.592  \pm 0.613$ & $47.386  \pm 0.000  (\pm0.000) $&    774   & $1.353  \pm 0.671$ & $47.244  \pm 0.000  (\pm0.000)$\\
     14.85   &    113  &  $1.592  \pm 0.613$ & $47.417  \pm 0.036  (\pm0.003) $&    774   & $1.353  \pm 0.671$ & $47.281  \pm 0.030  (\pm0.001)$\\
     14.90   &    113  &  $1.592  \pm 0.613$ & $47.454  \pm 0.067  (\pm0.006) $&    774   & $1.353  \pm 0.671$ & $47.328  \pm 0.051  (\pm0.002)$\\
     14.95   &    113  &  $1.592  \pm 0.613$ & $47.502  \pm 0.090  (\pm0.008) $&    774   & $1.353  \pm 0.671$ & $47.384  \pm 0.064  (\pm0.002)$\\
     15.00   &    113  &  $1.592  \pm 0.613$ & $47.546  \pm 0.108  (\pm0.010) $&    772   & $1.356  \pm 0.669$ & $47.432  \pm 0.077  (\pm0.003)$\\
     15.05   &    113  &  $1.592  \pm 0.613$ & $47.558  \pm 0.119  (\pm0.011) $&    752   & $1.383  \pm 0.657$ & $47.453  \pm 0.087  (\pm0.003)$\\
     15.10   &    110  &  $1.622  \pm 0.595$ & $47.549  \pm 0.131  (\pm0.013) $&    695   & $1.454  \pm 0.633$ & $47.455  \pm 0.097  (\pm0.004)$\\
     15.15   &    103  &  $1.685  \pm 0.560$ & $47.544  \pm 0.150  (\pm0.015) $&    599   & $1.574  \pm 0.600$ & $47.468  \pm 0.110  (\pm0.005)$\\
     15.20   &     90  &  $1.798  \pm 0.507$ & $47.550  \pm 0.179  (\pm0.019) $&    481   & $1.741  \pm 0.553$ & $47.491  \pm 0.128  (\pm0.006)$\\
     15.25   &     75  &  $1.935  \pm 0.439$ & $47.563  \pm 0.207  (\pm0.024) $&    386   & $1.891  \pm 0.515$ & $47.509  \pm 0.134  (\pm0.007)$\\
     15.30   &     59  &  $2.079  \pm 0.381$ & $47.589  \pm 0.207  (\pm0.027) $&    269   & $2.108  \pm 0.472$ & $47.532  \pm 0.157  (\pm0.010)$\\
     15.35   &     41  &  $2.241  \pm 0.345$ & $47.616  \pm 0.189  (\pm0.030) $&    173   & $2.350  \pm 0.422$ & $47.556  \pm 0.149  (\pm0.011)$\\
\hline
\end{tabular}
\end{minipage}
\end{table*}

\begin{table*}
 \centering
 \begin{minipage}{\textwidth}
  \centering
  \caption{Mean values of the spectral index $\alpha$ in the optical$-UV$ region, standard deviation and error on the mean, for the RL and RQ subsamples. Col.~6 reports the Kolmogorov-Smirnov probability that the RLQs and the RQQs (or RQQ matched sample) are drown from the same population.}\label{pertantiz}
\begin{tabular}{@{}c|cccc|cccc|c@{}}
\hline 
\hline
Samples &\# RLQ&$\alpha_{\rm RLQ}$&\# RQQ&$\alpha_{\rm RQQ}$&$P_{\rm KS}$\\
(1)&   (2)& (3)              & (4)  & (5) 	       & (6)\\
\hline
Total sample&113&$-0.55\pm0.41(\pm0.04)$&774&$-0.31\pm0.32(\pm0.01)$&$10^{-7}$\\
\hline
Matched samples&$''$&$''$&112&$-0.36\pm0.29(\pm0.03)$&$10^{-3}$\\
%&$''$&$''$&120&$-0.35\pm0.30(\pm0.03)$&$10^{-5}$\\
%&$''$&$''$&126&$-0.33\pm0.31(\pm0.03)$&$10^{-6}$\\
%&$''$&$''$&\ldots &\ldots &\ldots\\
%\hline
%Matched samples&$''$&$''$&119&$-0.38\pm0.29(\pm0.03)$&$10^{-2}$\\
%&$''$&$''$&120&$-0.36\pm0.32(\pm0.03)$&$10^{-3}$\\
%&$''$&$''$&126&$-0.34\pm0.35(\pm0.03)$&$10^{-4}$\\
%&$''$&$''$&126&$-0.36\pm0.29(\pm0.03)$&$10^{-3}$\\
%&$''$&$''$&107&$-0.36\pm0.29(\pm0.03)$&$10^{-3}$\\
%&$''$&$''$&103&$-0.36\pm0.29(\pm0.03)$&$10^{-3}$\\
%&$''$&$''$&102&$-0.36\pm0.29(\pm0.03)$&$10^{-3}$\\
%&$''$&$''$&105&$-0.36\pm0.29(\pm0.03)$&$10^{-3}$\\
%&$''$&$''$&105&$-0.36\pm0.29(\pm0.03)$&$10^{-3}$\\
%&$''$&$''$&105&$-0.36\pm0.29(\pm0.03)$&$10^{-3}$\\
%&$''$&$''$&104&$-0.36\pm0.29(\pm0.03)$&$10^{-3}$\\
%&$''$&$''$&\ldots &\ldots &\ldots\\
\hline
$z<0.8$   &12&$-0.64\pm0.55(\pm0.16)$&181&$-0.21\pm0.37(\pm0.03)$&$10^{-2}$\\
$0.8<z<1.2$&23&$-0.41\pm0.34(\pm0.07)$&188&$-0.26\pm0.31(\pm0.02)$&$10^{-1}$\\
$1.2<z<1.6$&22&$-0.66\pm0.50(\pm0.11)$&157&$-0.34\pm0.30(\pm0.02)$&$10^{-4}$\\
$1.6<z<2.0$&28&$-0.56\pm0.40(\pm0.08)$&117&$-0.37\pm0.27(\pm0.03)$&$10^{-2}$\\
$z>2.0$&28&$-0.51\pm0.34(\pm0.07)$&131&$-0.42\pm0.28(\pm0.02)$&$10^{-1}$\\
\hline
\end{tabular}
\end{minipage}
\end{table*}

%\subsection{Comparison between the SED of RLQs and RQQ matched samples}
%\label{SED_comparison}
%It is apparent from Fig.~\ref{figall} that RLQs are on average redder and more luminous than RQQs. 
Since we are interested in the study of the shape of the SED with reference to the radio power, independently of any other parameter, we now construct a number of RQQ samples matched in absolute magnitude to the RLQs\footnote{An RQQ with a given luminosity $L$ is extracted from the RQQ population with probability proportional to the luminosity density of the RLQs% $P_L(L)$
.}. We verify that the RQQ matched samples automatically result well matched to the RLQs also in redshift.

Table \ref{pertantiz} reports the average results of 10 random matched samples. The SEDs of the RQQ matched samples are on average slightly  softer than those of the entire RQQ population ($\Delta\alpha=0.05$),
%$\langle\alpha_{\rm RQQ\,m.s.}\rangle=-0.36$; $\langle\alpha_{\rm RQQ}\rangle=-0.31$), 
however the
%compare the RLQ population with a number of RQQ matched samples. , as described in section \ref{SEC_matched_sample}. The RQQ matched samples have the same distributions in redshift and magnitude of the RLQs. %a sample of RQQs well matched with RLQs both in redshift and luminosity distributions. !!!!!!!!!!!!!!!!!!!!!!!!!!!!!!!!!!!!!!!!!!!!!select from the radio quiet population a subsample of objects which have the same distributions in redshift and magnitude of the RLQs, in order to perform a statistical comparison in the $\alpha$ parameter between two well matched samples. The test is repeated for a number of ..............................................................................................
%Again, a 
Kolmogorov-Smirnov test indicates that the probability of RLQs and all the RQQ matched samples to be drawn from the same underlying population is still negligible ($P\sim 10^{-3}$; the value is greater than before as the sample has been drastically reduced), and the difference of the mean spectral indices is again $\sim0.2$. This is consistent with the fact that the whole RQQ population and an RQQ matched sample have identical distributions in $\alpha$ ($P\sim 90\%$) and it is a strong indication that the radio loud and radio quiet populations are intrinsically different, in the sense that radio loud objects are systematically redder than radio quiet quasars independently of the average luminosity or cosmic time. 

%The same result is obtained when comparing RLQs with the whole RQQs population: this can be seen also from figure \ref{sed_pochipunti}, in which it is apparent that RLQs are on average more luminous (consistently with Figure \ref{Mdistrib}) and, again, that the underlying power law is steeper for RQQs. Table \ref{pertantiz} reports the mean values of $\alpha$, the standard deviation and the error on the mean value for the RL and RQ subsamples. %Note that not only RLQs are significantly redder than RQQs: (NON CHIARO) they have also a larger dispersion on the spectral index. This means that redder objects are likely to be radio loud, while among bluer objects both RLQs and RQQs can be found. Note however that the difference in the standard deviation of the two subsamples may result from a bias introduced in the sample definition; indeed this is not the case for the observed difference between the average spectral indices (see section (\ref{SED_bias}) for a discussion on this point).

Finally we compare RLQs and RQQs in different redshift bins
%\footnote{The division in redshift bins is such that each bin contains approximately the same number of RL and RQ objects.} 
(see again Table \ref{pertantiz}), and this test shows that the color difference between the RL and RQ populations is apparent in all the redshift ranges. %The results of this test are also reported in table \ref{pertantiz}.

\section{Summary and discussion}
\label{sum_disc}
The aim of the present paper is the study of the continuum emission in the NIR to UV region of a quasar sample, in order to compare and contrast the SEDs of the RLQ and RQQ subsamples. We selected a sample of quasars with both SLOAN and 2MASS detection, to study a spectral range from the NIR (for low-$z$ objects) to the UV (for high-$z$ QSOs). 
The sample consists of 887 objects, of which 113 RLQs and 774 RQQs.

For each subsample we constructed the mean SED and evaluated the average spectral index in the NIR to UV region. 
The slope of the underlying power-laws is significantly different: the spectral indices of the two population are statistically different at more than 99\% confidence, both for the whole sample and dividing the sample in redshift bins. The difference is present also considering samples well matched in absolute magnitude and redshift. 

If the blue bump is due to the superposition of black body emissions from an accretion disc, then the color difference between RLQs and RQQs should be interpreted in terms of different mean temperatures, in the sense that RQQs are hotter.

%It has been suggested (Francis et al., ...) that the origin of the color difference between RLQs and RQQs may be linked to an enhanced synchrotron component due to the relativistic beaming of the jets of radio loud objects; note anyway that an interpretation in terms of synchrotron emission does not justify the difference observed in the optical-$UV$ region. 

\begin{figure}
\includegraphics[width=0.45\textwidth]{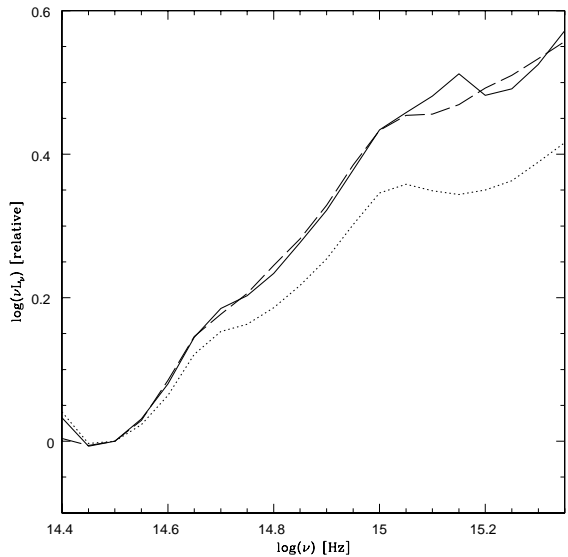}
\caption[]{Comparison of the SED of RQQs (dashed line) to the SED of RLQs (dotted line). When RLQs are corrected for an additional dust extinction ($\Delta A_{V}=0.16$mag, solid line), no difference is apparent.}
\label{dust}
\end{figure}

We first examine the possibility that the difference in the SEDs of RLQs and RQQs is due to an enhanced dust extinction in radio loud objects, as it has been suggested e.g.~by Francis et al.~(2000). In Fig. \ref{dust} we compare the SED of RQQs to the SED of RLQs, when RLQs are additionally corrected for dust extinction with respect to RQQs: the offset between the SEDs of the two population is minimized adopting $\Delta A_{V}=0.16$mag. The differences are now completely negligible, supporting the hypothesis that RL objects are more subject to dust extinction than RQQs.
The problem would be reconducted to understand why RLQs are more extincted than RQQs. {\it A priori} this could be related to a difference in the inclination angle distribution. %, for which however no immediate explanation appears available. 
Alternatively, it could be that the conditions of dust production are related to those justifying large radio emission. %, but again no explanation is apparent. 

We then focus on the possibility that the difference of disc temperature of RLQs and RQQs is real. Since the temperature of the inner disc scales as $M_{\rm BH}^{\,\,\,-0.25}$ (e.g.~Shakura \& Sunyaev 1973), the difference may be attributed to the fact that the black holes of radio loud quasars are supposedly more massive (e.g.~Dunlop et al.~2003; Falomo et al.~2004; Labita et al.~2007). %However, the mass difference is generally inferred from the larger luminosity of RLQs, and we have shown that the color difference is found also in luminosity matched samples.

%Obviously it is possible that the continua of RLQs and RQQs are intrinsically different, with RLQs being cooler,
%The spectral range on which we observe a difference in the shapes of the SEDs of RLQs and RQQs is usually said to be dominated by thermal emission from an accretion disk, and we may wonder if that could be related to the fact that BHs are suspected to be somewhat more massive in RLQs, since the temperature of gas and radiation in linear accretion disks scales with the BH mass as $T\propto M_{\rm BH}^{\,\,\,-\frac{1}{4}}$ (Shakura \& Sunyaev 1973). This would explain why RLQs are cooler (redder) than RQQs, but note that the color difference is apparent also in the comparison between RLQs and the RQQ matched samples, which have the same distribution in luminosity and then, probably, in BH mass. 
The color difference may be linked to the BH spin (Stawarz, Sikora \& Lasota 2007), as radio emission is usually ascribed to a faster spinning. However, spinning BHs are expected to have a shorter last stable orbit radius, and then a hotter disc, inconsistently with our results. %So the possibility that the thermal contribution from the accretion disk is softer in RLQs can be ruled out. 

Obviously it is possible that the non--thermal (power law) continua of RLQs and RQQs are intrinsically different. Supposing that the non--thermal component accounts for 80\% at $10^{14.5}$Hz, while at $10^{15.35}$Hz (in the Big Blue Bump) the thermal component is dominant and accounts for 80\%, our data are consistent with a picture where the underlying power--laws of RLQs and RQQs have $\alpha_{\rm RLQ}=-1.2$ and $\alpha_{\rm RQQ}=-1.0$ respectively, whereas the thermal bumps are indistinguishable. The problem would be reconducted to understand why the non--thermal continuum of RLQs is softer than RQQs. %, but again no obvious explanation can be invoked. 
It has been suggested (i.e.~Francis et al.~2000) that in radio loud samples there is a significant chance of synchrotron contamination of the rest-frame R--band nuclear luminosities, due to the presence of the relativistic jets. This effect explains the red color of QSOs in  high-frequency selected radio samples (i.e. the PKS survey, e.g.~Francis et al.~2001), that suffer from a bias towards pole-on radio sources which are
relativistically boosted above the survey flux limit, but no obvious explanation can be invoked to justify the color difference in our sample.

%Fig.~\ref{final_LQ} shows the selection criteria $g<18.9$ and the distribution of the RLQs and RQQs in the $M_g$ vs. $z$ plane, while Fig.~\ref{LQ_z_hist} shows the distribution in redshift of the RLQs and RQQs\footnote{Note that the Kolmogorov-Smirnov probability that the reduced sample is randomly drawn from the SLOAN catalogue for what concerns the variable $u-r$ is nearly 0. This is probably due to the fact that the selection limit in the apparent magnitude introduces a cut in redshift, while the observed $u-r$ depends on $z$. The same probability evaluated for smaller redshift bins increases with reducing the $z$ range.
\section*{Acknowledgements}
We are grateful to Dr.~M.~Strauss for constructive criticism, and to Dr.~D.~Bettoni for help in the use of SLOAN archives.
This work was partially supported by PRIN 2005/32. 

\begin{footnotesize}

\end{footnotesize}

\label{lastpage}
\end{document}